\documentclass[aps,showpacs,preprintnumbers,amsmath,amssymb]{revtex4}

\usepackage{float}
\usepackage{graphics,epsfig}
\usepackage{graphicx}
\usepackage{dcolumn}
\usepackage{bm}
\usepackage{slashbox}
\usepackage{multirow}

\begin{document}
\baselineskip=0.8 cm

\title{{\bf  Non-equilibrium condensation process in holographic superconductor with nonlinear electrodynamics}}

\author{Yunqi Liu}
\email{liuyunqi@hust.edu.cn}
\author{Yungui Gong}
\email{yggong@mail.hust.edu.cn}
\affiliation{School of Physics, Huazhong University of Science and Technology, Wuhan, Hubei 430074, China}
\author{Bin Wang}
\email{wang_b@sjtu.edu.cn}
\affiliation{IFSA Collaborative Innovation Center, Department of Physics and Astronomy, Shanghai Jiao Tong University, Shanghai 200240, China}

\date{\today}
\begin{abstract}
\baselineskip=0.6 cm
We study the non-equilibrium condensation process in a holographic superconductor with nonlinear corrections to the U(1) gauge field. We start with an asymptotic Anti-de-Sitter(AdS) black hole against a complex scalar perturbation at the initial time, and solve the dynamics of the gravitational systems in the bulk. When the black hole temperature $T$ is smaller than a critical value $T_c$, the scalar perturbation grows exponentially till saturation, the final state of spacetime approaches to a hairy black hole. In the bulk theory, we find the clue of the influence of nonlinear corrections in the gauge filed on the process of the scalar field condensation. We show that the bulk dynamics in the non-equilibrium process is completely consistent with the observations on the boundary order parameter. Furthermore we examine the time evolution of horizons in the bulk non-equilibrium transformation process from the bald AdS black hole to the hairy AdS black hole. Both the evolution of apparent and event horizons show that the original AdS black hole configuration requires more time to finish the transformation to become a hairy black hole if there is nonlinear correction to the electromagnetic field. We generalize our non-equilibrium discussions to the holographic entanglement entropy and find that the holographic entanglement entropy can give us further understanding of the influence of the nonlinearity in the gauge field on the scalar condensation.

\end{abstract}
\maketitle

\vspace*{0.2cm}

\baselineskip=0.8 cm

\section{Introduction}

The AdS/CFT correspondence has been proved as a unique approach to study the strongly coupled field theories \cite{Maldacena,S.S.Gubser-1,E.Witten}. Recently, there appeared a new direction in applying this correspondence to shed some lights in condensed matter physics, with the hope that the strongly coupled systems can be computationally tractable and conceptually more transparent (for a review, see \cite{S.A. Hartnoll,G.T. Horowitz-1,CaiLiLiYang} and references therein). The simplest example of the duality between the superconductor and gravity theory was first formulated in \cite{3H}. In the gravity side one has Einstein-Maxwell-charged scalar theory with negative cosmological constant, which usually has static Reissner-Nordstrom-AdS (RN-AdS) black hole solutions with vanishing scalar hair. However in the low temperature, because of the coupling between the Maxwell field and the charged scalar field, the RN-AdS black hole is expected to become unstable and finally the black hole will have the scalar hair breaking the $U(1)$-gauge symmetry spontaneously. The RN-AdS and hairy black holes are remarkably identified as the normal and superconducting phases of a superconductor realized in the dual field theory on the AdS boundary. This striking connection between condensed matter and gravitational physics is expected to be a hopeful approach to disclose the properties of strongly correlated electron systems. Most available studies on this topic were concentrated on the static AdS black hole backgrounds dual to the superconductors reflected by the dual operators on the AdS boundary with equilibrium or nearly equilibrium dynamics \cite{S.S. Gubser}-\cite{CD}.

As gravitational theorists, one of our motivations of seeking a dual description of superconductivity, which is below our usual energy bandwidth, is to understand more on the black hole physics, for example how the scalar hair attaches to the bald AdS black hole to make the black hole become hairy step by step, what conditions stimulate or prevent the condensation of the scalar hair onto the AdS black hole, what characteristics of the transformation between the AdS bald black hole and the hairy one are, what spacetime properties are for the developed hairy AdS black holes, etc. To answer these questions, we cannot just look at the dual operator on the AdS boundary in an equilibrium system, we need to concentrate on the non-equilibrium dynamics in the bulk  gravitational configurations to disclose the process of addressing the scalar hair onto the bald AdS black hole. In \cite{Murata}, the non-equilibrium process in the bulk Einstein-Maxwell-scalar background has been studied. It was observed that during the time evolution, the initial scalar perturbation in the bulk spacetime first moves towards the AdS boundary, it is then bounced back by the AdS wall, grows and moves towards the bald black hole. Subsequently this perturbation falls onto the bald black hole and accumulates there and finally makes the bald black hole become hairy. The time evolutions of the event and apparent horizons of the bulk black hole in the AdS bulk were disclosed in \cite{Murata}. Employing the AdS/CFT duality, the authors related the observed properties in the AdS bulk to the boundary CFT, and further they displayed the time dependence of the superconducting order parameter in the holographic superconductor. The attempt in the non-equilibrium approach has been generalized to the study of the dynamical phase diagram corresponding to quantum quenches of source field where the bulk black hole perturbation spectrum has been investigated in relating to the distinct behaviors on the boundary CFT \cite{BhaseenWiseman}. More extensions of the non-equilibrium physics in describing the transformation of an unstable anisotropic black hole without scalar hair into a stable hairy black hole, which relates to the  holographic superconductor with anisotropy, have been done in \cite{Bai}. Further works on the far-from-equilibrium dynamics of holographic superconductor including periodic driven system \cite{LiTianZhang}, quantum quench of a superconducting AdS soliton \cite{GaoGarciaZeng},  holographic thermalization \cite{GarciaGarciaZeng,Buchel} and the formation of topological defects \cite{JulianSonner,CheslerLiu} have also been extensively studied.

Motivated by the application of the Mermin-Wagner theorem to the holographic superconductors, there have been a lot of interest in generalizing the Einstein gravity background in the holographic superconductor to the gravity including the curvature correction. By examining the charged scalar field together with a Maxwell field in the Gauss-Bonnet-AdS black hole background \cite{R. Gregory,K. Maeda,BarclayGregoryKannoSutcliffe,Siani,JingWangPanChen,GregorJPCS,BarclayJHEP1110,CaiNieZhang,LiCaiZhang,LiuPanWangCai,Kanno,Gangopadhyaya}, it has been observed that the higher curvature correction makes the holographic superconductor harder to form. Besides the curvature correction to the gravity, it is also interesting to investigate the high-order correction related to the gauge field. In the study of the ratio of shear viscosity to entropy density, it was found that the higher derivative correction to the gravity has different effect from that of the higher derivative correction to the gauge field \cite{CaiSun}. This motivates people to study the effect of higher order corrected Maxwell field on the scalar condensation and compare with the effect brought by the correction in the gravity \cite{PanJingWang}. 

Since Heisenberg and Euler \cite{Heisenberg} noted that quantum electrodynamics predicts that the electromagnetic field behaves nonlinearly through the presence of virtual charged particles, the nonlinear electrodynamics has been studied a lot for many years \cite{Heisenberg,GrossSloan,Born,Gibbons,OliveraOlea,Hoffmann,Oliveira}. In the low-energy limit of heterotic string theory, the higher-order correction term appears also in the Maxwell gauge field \cite{GrossSloan}. Born-Infeld electrodynamics \cite{Born} is the only possible nonlinear version of electrodynamics which is invariant under electromagnetic duality transformations \cite{Gibbons}. The nonlinear electromagnetic fields in logarithmic form appears in the description of vacuum polarization effects. The form was obtained as exact 1-loop corrections for electrons in an uniform electromagnetic field background by Euler and Heisenberg \cite{Heisenberg}. In \cite{Soleng}, the author discussed a simple example of a Born-Infeld-like Lagrangian with a logarithmic term which can be added as a correction to the original Born-Infeld one. Replacing the Maxwell field with the nonlinear electromagnetic field, it was disclosed from the behavior of the boundary operator in the CFT that the nonlinearity in Born-Infeld and the logarithmic form fields can make holographic superconductor difficult to form \cite{J.L. Jing,ZhaoJingChen}. In the corresponding bulk spacetime, it was observed that the nonlinearity of the electromagnetic field can hinder the growing mode to appear in the scalar perturbation \cite{liuandwang}, which is consistent with the phenomenon on the boundary CFT. Other works on holographic dual models in the presence of nonlinear corrections to the Maxwell electrodynamics can also be found in \cite{WuCaoKuang,Lee,BaiGaoQi}. These discussions were concentrated on the equilibrium state both in the bulk and boundary. To understand more on how these two kinds of nonlinear corrections in the electromagnetic fields influence the scalar condensation in the holographic superconductor, we can not only concentrate on the equilibrium state but also study the non-equilibrium condensation process in holographic superconductor with nonlinear electrodynamics. This can help us see closely how the nonlinearity in the electrodynamics influences the condensation of the scalar hair in the process of the black hole evolution and transformation, it also can aid us to see how the nonlinearity affect the phase transition in the boundary system . 

In this paper we will concentrate more on the non-equilibrium bulk spacetime. The organization of the paper is the following: In Section II, we will  introduce the holographic superconductor model with nonlinear electrodynamics. We will give equations of motion and the initial and boundary conditions to solve them. In Section III, we will show the numerical results in the non-equilibrium bulk system. We will present that the condensation process of the scalar field depends on the strength of the nonlinearity in the electromagnetic field. We will describe the time evolution of the event and apparent horizons of the black hole in the bulk. Furthermore, we will display the time evolution of the holographic entanglement entropy and its dependence on the nonlinearity of the electromagnetic field. We will summarize and discuss our results in the last section.

\section{model}
The action describing a gauge field coupled with a scalar field  with a negative
cosmological constant in the 4D spacetime reads
\begin{eqnarray}\label{action}
S=\int d^4
x\sqrt{-g}\left[R+\frac{6}{l^2}+L(F)-|\nabla\Psi-i q A\Psi
|^2-m^2|\Psi|^2 \right]
\end{eqnarray}
where $l$ is the AdS radius which is set to unity hereafter,  $q$ is the coupling factor between the gauge field and the complex scalar field, $m$ is the mass of the scalar field. $L(F)$ is the Lagrangian density of the nonlinear electromagnetic field. In this work, we will investigate the nonlinear electromagnetic fields in two different forms, the Born-Infeld form (BINE) and the Logarithmic form (LNE), 
$$L(F)=\left\{
\begin{aligned}
&\frac{1}{b^2}\left(1-\sqrt{1+\frac{1}{2}b^2 F^2}\right),~~~~~\text{BINE},\\
& -\frac{2}{b^2}\text{Ln}\left(1+\frac{1}{8}b^2 F^2 \right),~~~~~~~\text{LNE},
\end{aligned}
\right.$$
where $F^2=F^{\mu\nu}F_{\mu\nu}$, and $b$ is the nonlinear parameter.
As $b\rightarrow0$, both kinds of the nonlinear electromagnetic fields 
reduce to the Maxwell field,  $L(F)\rightarrow-F^2/4$, and the model action
(\ref{action}) reduces to that in \cite{Murata}.

We take the ansatz of the background as the ingoing Eddingthon-Finkelstein coordinates,
\begin{eqnarray}\label{ansatz1}
ds^2=-\frac{1}{z^2}[f(v,z)dv^2+2 dv dz]+ \phi^2(v,z)(dx^2+dy^2).
\end{eqnarray}
where $v$ is the ingoing Eddington-Finkelstein time coordinate, $z$ is
the radial direction in the bulk and the AdS boundary is located at $z=0$.
With the axial gauge, the ansatz of the gauge field and scalar field can be written as,
\begin{eqnarray}\label{ansatz2}
A=(\alpha(v,z),0,0,0), ~~~~~~~~~~
\Psi=\psi(v,z).
\end{eqnarray}
There are still the following residual gauge symmetries for the anstaz which will be fixed by the boundary conditions
\begin{eqnarray}\label{residual}
\frac{1}{z}\rightarrow \frac{1}{z}+g(v),~~\alpha\rightarrow\alpha+\partial_v \theta(v),~~\psi\rightarrow e^{i q\theta(v)}\psi.
\end{eqnarray}
From the action and the ansatz we can derive the equations of motion as follows,
\begin{eqnarray}\label{EOMS}
&&(\phi D\phi)'+\frac{\phi^2}{4z^2}(6-m^2 |\psi|^2+L(F)-\frac{1}{2}z^4 F_2\alpha'^2)=0,\label{EOMS1}\\
&&D\psi'+\frac{\phi'}{\phi}D\psi+\frac{D\phi}{\phi}\psi'+\frac{1}{2}i q \alpha' \psi+\frac{m^2}{2z^2} \psi=0,\label{EOMS2}\\
&&(z^2(\frac{f}{z^2})')'-z^2 \alpha'^2 F_2+4 \frac{\phi'}{\phi^2}D\phi-(\psi'^{*}D\psi+\psi'D\psi^{*})=0,\label{EOMS3}\\
&&2z^2 F_2 (D\alpha)'+z^4(z^{-2}fF_2)'\alpha'+4z^2 F_2\frac{D\phi}{\phi}\alpha'\nonumber\\&&+2z^2 DF_2
\alpha'-z^2 f F_2'\alpha'-2iq(\psi D\psi^*-\psi^*D\psi)=0,\label{EOMS4}\\
&&D^2\phi+\frac{1}{2}(\frac{f}{z^2})'D\phi+\frac{\phi}{2}|D\psi|^2=0,\label{cons1}\\
&&(z^2 \phi')'+\frac{z^2}{2} \phi |\psi'|^2=0,\label{cons2}\\
&&(z^2 F_2 \alpha')'+2z^2F_2\alpha' \frac{\phi'}{\phi}+i q(\psi'^*\psi-\psi^*\psi')=0\label{cons3}
\end{eqnarray}
where the prime denotes the derivative with respect to $z$, i.e. $'=\partial_z$, and the derivative operator $D$ is defined as
\begin{eqnarray}
D\phi&=&\partial_v \phi-f \partial_z \phi/2,~~~~~~~~~~~~~~~~~~~D^2 \phi=\partial_v (D\phi)-f \partial_z (D\phi)/2,\nonumber\\
D\psi&=&\partial_v \psi-f \partial_z \psi/2-iq\alpha\psi, ~~~~~~~~D\psi^*=\partial_v \psi^*-f \partial_z \psi^*/2+iq\alpha\psi^*,\nonumber\\
D\alpha&=&\partial_v\alpha-f\partial_z \alpha/2,~~~~~~~~~~~~~~~~~~DF_2=\partial_vF_2-f\partial_zF_2/2.\nonumber
\end{eqnarray}
The function $F_2$ is defined as $F_2=(1+b^2F^2/2)^{-1/2}$ for BINE model,
while for the LNE model it reads $F_2=(1+b^2 F^2/8)^{-1} $. In the limit $b\rightarrow0$,
$F_2\rightarrow1$,  the system and the equations of motion  go back to those described in \cite{Murata}. In our numerical computation, we will mainly consider the first four equations, Eqs. (\ref{EOMS1})$-$(\ref{EOMS4}), describing the evolutions of the variables $f,\phi,\alpha$, and $\psi$. We will treat the remaining three equations as constraints.

\subsection{the normal phase}
While the scalar field is null, the solution to the equations of motion is a static AdS black hole solution. As the nonlinear parameter $b$ approaches zero, $b\rightarrow0$, the nonlinear electromagnetic field goes back to the Maxwell field, the background is nothing but the Reisser-Nordstr\"{o}m-AdS black hole, the solution is given by
\begin{eqnarray}\label{MAXBH}
\phi=\frac{1}{z},~~~\psi=0,~~~\alpha=Qz,~~~f=1-2M z^3+\frac{1}{4}Q^2 z^4.
\end{eqnarray}

While $b\neq0$, one can solve the equations of motion directly, the solutions are given as follows:

$(i)$ In the case of BINE
\begin{eqnarray}\label{BINEBH}
\phi&=&\frac{1}{z},~~~\psi=0,~~~\alpha=Q z \,_2F_1[\frac{1}{4},\frac{1}{2},\frac{5}{4},-b^2Q^2z^4] \nonumber\\
f&=&1-2Mz^3+\frac{1}{6b^2}(1-\sqrt{1+b^2Q^2z^4})+\frac{Q^2z^4}{3}\,_2F_1[\frac{1}{4},\frac{1}{2},\frac{5}{4},-b^2Q^2z^4] .
\end{eqnarray}

$(ii)$ In the case of LNE
\begin{eqnarray}\label{LNEBH}
\phi&=&\frac{1}{z},~~~\psi=0,~~~\alpha=\frac{2}{3 b^2 Q z^3}(1-\sqrt{1+b^2 Q^2 z^4})+\frac{4}{3}Q z\,_2F_1[\frac{1}{4},\frac{1}{2},\frac{5}{4},-b^2Q^2z^4],\nonumber\\
f&=&1-2 M z^3+\frac{5}{9 b^2}(1-\sqrt{1+b^2 Q^2 z^4})\nonumber\\&+&\frac{4 Q^2 z^4}{9}
\,_2F_1[\frac{1}{4},\frac{1}{2},\frac{5}{4},-b^2Q^2z^4]-\frac{1}{3b^2}\text{Ln} (\frac{2(\sqrt{1+b^2 Q^2 z^4}-1)}{b^2 Q^2 z^4}).
\end{eqnarray}
where $\,_2F_1$ represents a hypergeometric function. $Q$ and $M$ are proportional to the charge and the mass in the field theory, respectively. The spacetime described in Eq. (\ref{BINEBH}) is nothing but the Born-Infeld-AdS black hole \cite{DeyCai} and the spacetime presented in Eq. (\ref{LNEBH}) is the same as that in \cite{Soleng}.

For these static black holes, the Hawking temperature is given by
\begin{eqnarray}
T=-\frac{1}{4 \pi}\frac{df}{dz}|_{z=z_h},
\end{eqnarray}
$z_h$ is the radius of black hole horizon defined by $f(z_h)=0$. These asymptotic AdS black holes are unstable when the temperature of the black hole $T$ is smaller than a certain critical value $T_c$. The $U(1)$ gauge symmetry will be spontaneously broken due to the condensation of the complex scalar field.

\subsection{Boundary and initial conditions}

We set the mass of the complex field as $m^2=-2l^{-2}$. The expansion
of the variables near AdS boundary behave as
\begin{eqnarray}\label{asymp}
f(v,z)&=&1+f_1(v)z+f_2(v)z^2+f_3(v)z^3+\ldots,\nonumber\\
\phi(v,z)&=&1/z+\phi_0(v)+\phi_1(v)z+\phi_2(v)z^2+\phi_3(v)z^3+\ldots,\nonumber\\
\psi(v,z)&=&\psi_1(v)z+\psi_2(v)z^2++\psi_3(v)z^3\ldots,\nonumber\\
\alpha(v,z)&=&\alpha_0(v)+\alpha_1(v)z+\alpha_2(v)z^2+\alpha_3(v)z^3\ldots.
\end{eqnarray}
With the residual symmetries Eq. (\ref{residual}), we can choose
both $\phi_0(v)$ and $\alpha_0(v)$ to be zero.
Substituting Eq. (\ref{asymp}) into Eqs. (\ref{EOMS1}-\ref{EOMS4}), solving them we can obtain the coefficients as
the functions of $\psi_1(v)$ and $\psi_2(v)$,
\begin{eqnarray}\label{asymp-res1}
f_1&=&0,~~~~~~~~f_2=-\frac{1}{2}\psi_1 \psi_1^*,~~~~~~~~\phi_1=-\frac{1}{4}|\psi_1|^2,\nonumber\\
\phi_2&=&-\frac{1}{6}(\psi_1\psi_2^*+\psi_1^*\psi_2),~~~~~~~~~~~~\phi_3=-\frac{1}{6}|\psi_2|^2-\frac{11}{96}|\psi_1|^4
-\frac{1}{8}(\psi_1\dot{\psi_2}^*+\psi_1^*\psi_2),\nonumber\\
\alpha_2&=&-\frac{1}{2}iq(-\psi_2\psi_1^*+\psi_1\psi_2^*),~~~~~~~~\psi_3=-\frac{1}{2}iq\alpha_1\psi_1+\dot{\psi_2}+\frac{1}{2}\psi_1^2\psi_1^*,
\end{eqnarray}
together with other two formulas,
\begin{eqnarray}\label{asymp-res2}
\dot{f_3}=\frac{1}{2}(\psi_1^*\ddot{\psi_1}+\psi_1\ddot{\psi_1^*}-\psi_1\dot{\psi_2^*}-\psi_1^*\dot{\psi_2}),
~~~~\dot{\alpha_3}=-iq(\psi_1\psi_2^*-\psi_2\psi_1^*+\dot{\psi_1}\psi_1^*-\psi_1\dot{\psi_1}^*).
\end{eqnarray}
where the overdot denotes the derivative with respect to $v$, $\dot{}\equiv d/dv$. $\alpha_3$ and $f_3$ are related to the charge and mass density of the system, respectively. To solve the system numerically, we employ the Chebyshev Pseoduspectral representation in $z$-direction, and an implicit finite difference scheme in $v$-direction. We found that 20 points in $z$-direction are sufficient to assure reliable results. The data presented in this paper is for 30 grids. Convergence testing shows the errors will be less than one percent in the plotted quantities.

In our numerical process, we consider $\psi_1(v)$ as the source of the boundary field and set it to be null at the AdS boundary $z=0$. In Eq. (\ref{asymp-res2}), it is interesting to see that if one chooses $\psi_1=0$, $f_3$ and $\alpha_3$ are conserved quantities in the system, which means that the black hole charge and mass are constants in the phase transition process. On the initial moment we specify the black hole $M$ and $Q$ as fixed parameters since they are conserved quantities in our setting. Without loss of generality, using the following scaling symmetry,
\begin{eqnarray}\label{}
&&(v,z,x,y)\longrightarrow(kv,kz,kx,ky),\nonumber\\
&&F\longrightarrow F, ~~~~\phi\longrightarrow \phi/k,~~~~ \alpha\longrightarrow \alpha/k, ~~~~\psi\longrightarrow \psi,\nonumber\\
&&M\longrightarrow M/k^3 , ~~~~Q\longrightarrow Q/k^2 ,~~~~ T\longrightarrow T/k.\nonumber
\end{eqnarray}
we can set the initial black hole horizons radius to unity. With $f(1)=0$ we can get

$(i)$ In the case of BINE
\begin{eqnarray}\label{}
M=\frac{1}{2}[1+\frac{1}{6b^2}(1-\sqrt{1+b^2Q^2})+\frac{Q^2}{3}\,_2F_1[\frac{1}{4},\frac{1}{2},\frac{5}{4},-b^2Q^2]].\nonumber
\end{eqnarray}

$(ii)$ In the case of LNE
\begin{eqnarray}\label{}
M&=&\frac{1}{2}[1+\frac{5}{9 b^2}(1-\sqrt{1+b^2 Q^2})\nonumber\\&+&\frac{4 Q^2 z^4}{9}
\,_2F_1[\frac{1}{4},\frac{1}{2},\frac{5}{4},-b^2Q^2]-\frac{1}{3b^2}\text{Ln} (\frac{2(\sqrt{1+b^2 Q^2}-1)}{b^2 Q^2})].\nonumber
\end{eqnarray}
thus only the black hole charge $Q$ is left as the open initial variable to be assigned. As for the initial condition, the background is given by the normal charged black holes, described by Eqs. (\ref{MAXBH}-\ref{LNEBH}), with null scalar hair, along with a scalar perturbation in the following form,
\begin{eqnarray}\label{}
\psi(z)=z^{2}A\exp\left[-\frac{(z-z_m)^2}{2\delta^2}\right],
\end{eqnarray}
with $A=0.01, \delta=0.05$ and $z_m=0.3$.

\section{Numerical Results}
In this section, we will show the numerical results for the non-equilibrium process for the holographic superconductor with nonlinear electrodynamics. We will first exhibit the dynamics of the scalar field, including its condensation in the bulk and  the evolution of the order parameter on the AdS boundary. Besides, we will disclose the evolution of both the event and apparent horizons. Furthermore, we will give the evolution of the holographic entanglement entropy in this time-dependent background. Hereafter, we take $q=4$ in our numerical calculation.

\subsection{The Condensation of Scalar Field}

To calculate the critical temperature $T_c$, we use the shooting method the same as those in \cite{LiuPanWang,K. Maeda,LiuPanWangCai}. With this method, we can reproduce the phase transition temperature in \cite{Murata}. The critical temperatures of black holes with nonlinear electromagnetic field in the forms BINE and LNE are listed in Table. I. We see that for both nonlinear electromagnetic field models, the critical temperature $T_c$ decreases with the increase of the nonlinearity. This tells us that the nonlinear correction to the Maxwell gauge field hinders the formation of the hairy black hole, which also means nonlinearity make the boundary system harder to reach the superconducting phase. The qualitative behavior of $T_c$ is consistent with that disclosed in the static black hole backgrounds in \cite{ZhaoJingChen}.

\begin{table}[ht]
\begin{tabular}{|c|c|c|c|c|c|c|}
\hline
\multicolumn{2}{|c|}{$b$} & 0 &0.6&1.2& 1.5& 1.8\\ \hline
\multirow{2}{0.3cm}{$T_c$}
 & BINE & 0.2195& 0.2112 & 0.1879& 0.1720 & 0.1543 \\ \cline{2-7}
&LNE &0.2195 &0.2154 & 0.2044& 0.1971 & 0.1889\\ \hline
\end{tabular}
\caption{The critical temperature $T_c$ as a function of the nonlinear parameter $b$ for both black holes with BINE and LNE with the backreacting parameter $q=4$.}
\end{table}

In Fig.\ref{con0}, starting from the initial black hole temperature $T\simeq0.5 T_c$,  we depict the time evolution of the scalar field 
$q|\psi(v,z)|$ as functions of $vT_c$ and $zT_c$  for different strengths of the nonlinearity in the electromagnetic field. The general feature of the condensation process is similar to that reported in \cite{Murata}. The initial Gaussian perturbation runs to the AdS boundary, then is reflected into the bulk and keeps growing exponentially until the saturation is reached. In the left panel, the parameter $b$ is fixed as $b=0$. We can see that the scalar field  keeps on increasing exponentially within the time interval $0<vT_c \lesssim 5$, but starting from $vT_c \simeq 5$ the scalar field becomes saturated at the horizon. The saturation of the scalar hair at the black hole horizon marks that the gravitational system enters into the phase of a  static hairy black hole state. At the saturation moment,  the scalar field $q|\psi(v,z)|$ near the black hole horizon reaches $q|\psi(v,z)|\simeq 1.3$. In the middle panel, we plot the figure with the parameter $b=1.8$ for the BINE model, while in the right panel we exhibit the result for the LNE model with the same parameter $b=1.8$. Comparing the left panel with the middle and right panels, we see that with the increase of the nonlinearity in the electromagnetic field, the scalar field needs longer period to reach the saturation, $vT_c \simeq 6$ in the middle panel. Besides we see that for the electromagnetic field with stronger nonlinearity, the final stable hairy black hole has more scalar hair, higher value of $q|\psi(v,z)|$ (for example $q|\psi(v,z)|\simeq 1.8$ in the middle panel). 

\begin{figure}[ht]\label{con0}
\centering
\includegraphics[width=145pt]{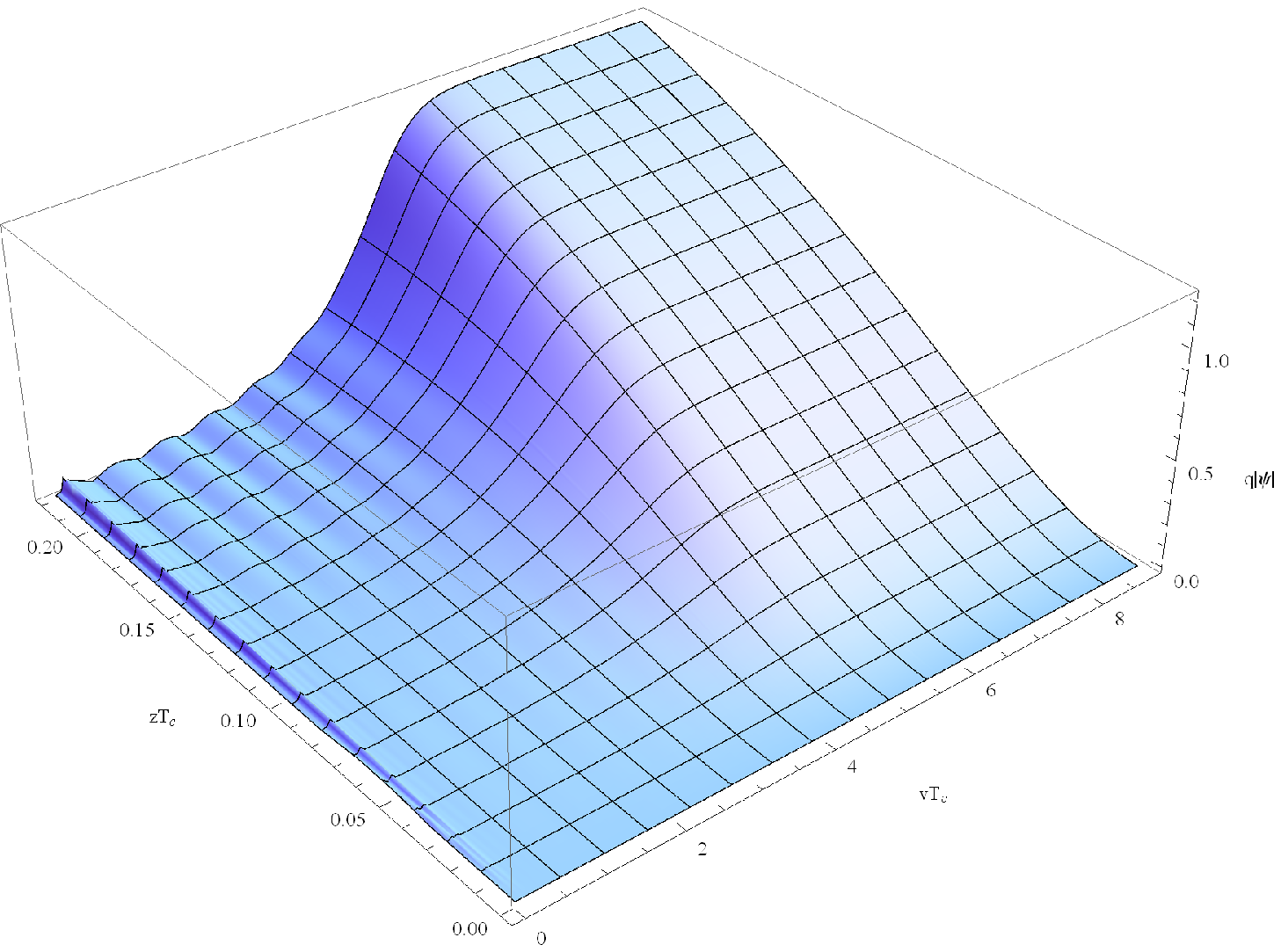}
\includegraphics[width=145pt]{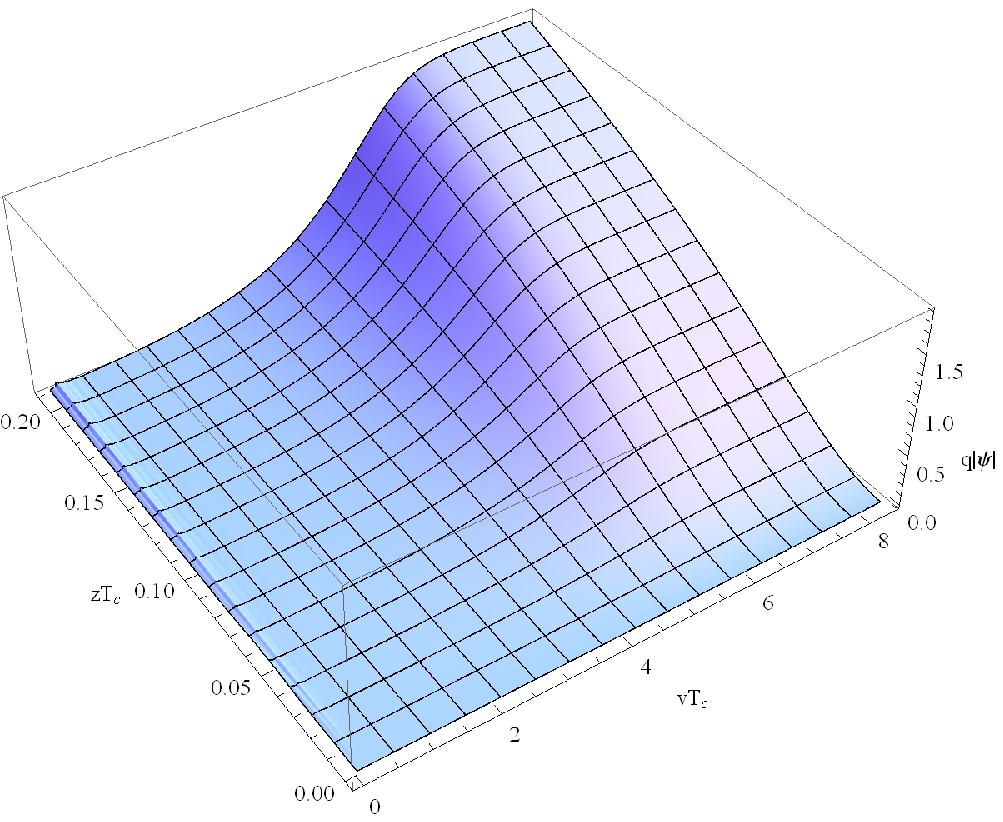}
\includegraphics[width=145pt]{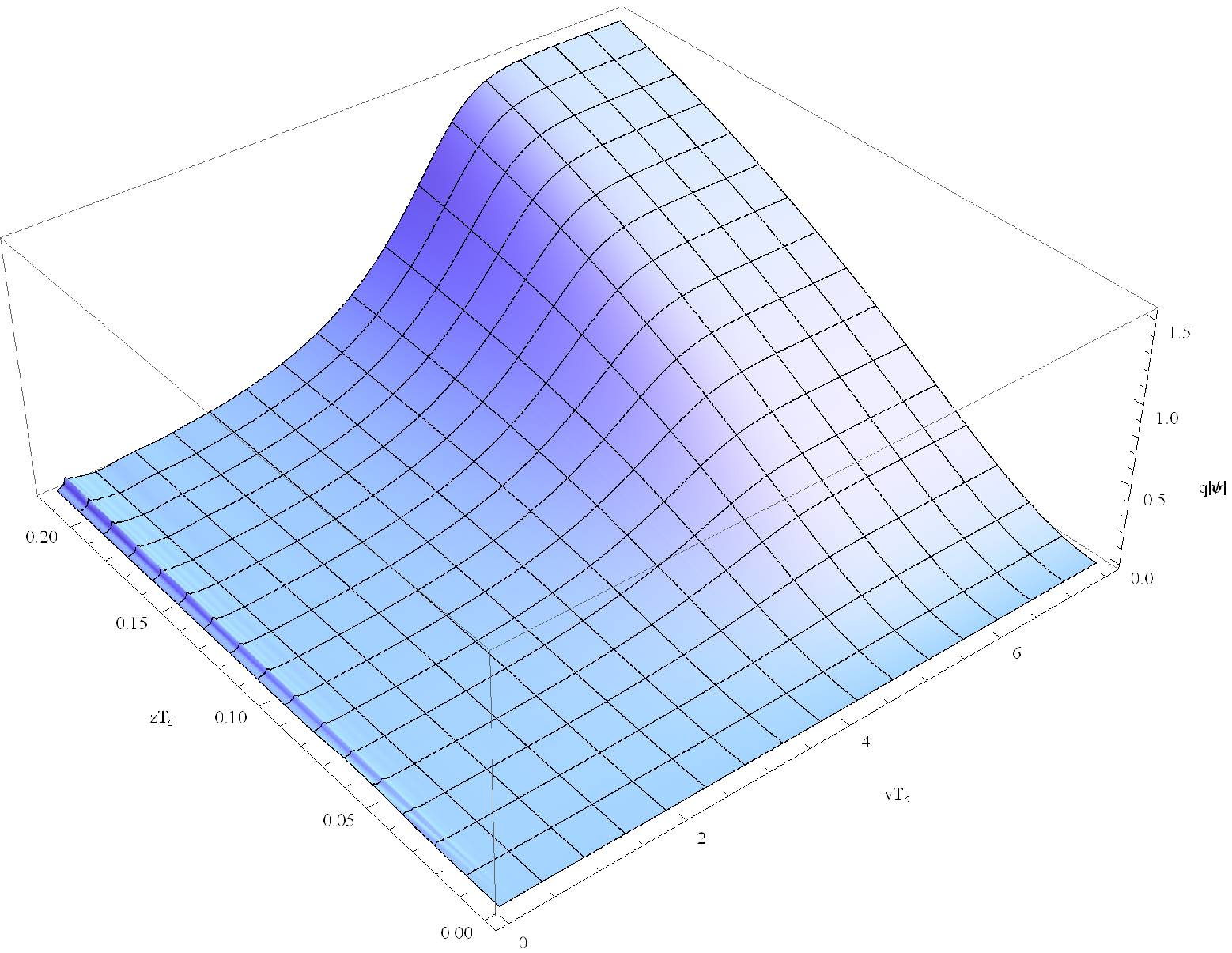}
\caption{\label{con0} The scalar field $|\psi(v,z)|$ as functions of $vT_c$ and $zT_c$, it grows exponentially till saturation. In the left panel the parameter $b=0$. The middle panel is for a system with BINE, and the right panel is for a system with LNE, $b$ is fixed as $b=1.8$.}
\end{figure}

\subsection{Evolution of the order parameter}\label{condensationpart}
In this subsection, we describe the dynamics of the order parameter
of the boundary theory. Following \cite{3H,Sean A. Hartnoll-3,BhaseenWiseman,Murata,Bai},
the order parameter is defined as
\begin{eqnarray}\label{}
\langle O_2(v)\rangle=\sqrt{2}\psi_2(v).
\end{eqnarray}
In Fig.\ref{operatorsL}, which displays the raw data, we show the numerical results with initial black hole temperature as $T\simeq0.5T_c$. It depicts the time-dependence of the order parameter $(q |\langle O_2(v) \rangle|)^{1/2}/T_c$ with $b=0,0.6,1.2,1.5$ and $1.8$ respectively, for both BINE and LNE models. From this figure, we see that no mater what the value of $b$ is, given enough time the order parameter will reach some certain values and will not increase any more, which implies that the system in the boundary CFT evolves into an equilibrium phase, i.e. the superconducting phase. The behaviors displayed in Fig. \ref{operatorsL} on the order parameters of the CFT on the boundary support the properties of the scalar hair evolutions disclosed in the bulk, for the model with stronger nonlinearity in the electromagnetic field, it costs more time for the order parameter to approach the saturation, the boundary is more difficult to arrive at the superconducting phase state. Fig.\ref{operatorsTc} shows the curves normalized by $T_c$.  With respect to Fig.\ref{operatorsL}, along the vertical direction the effect of $T_c$ is to increase the height of these curves, but it doesn't differ the order of these curves from bottom to top, also the properties of the curves. In horizontal direction, the effect of $T_c$ is to decrease the time period of non-equilibrium process in the boundary system, but the time interval with a bigger $b$ is still longer than that with a smaller $b$.

\begin{figure}[ht]\label{operatorsL}
\centering
\includegraphics[width=200pt]{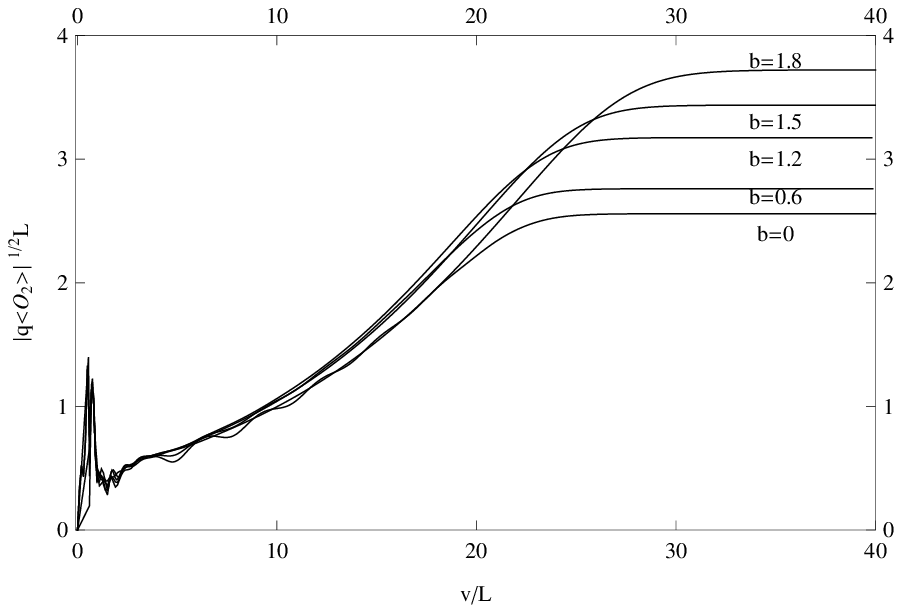}	\includegraphics[width=200pt]{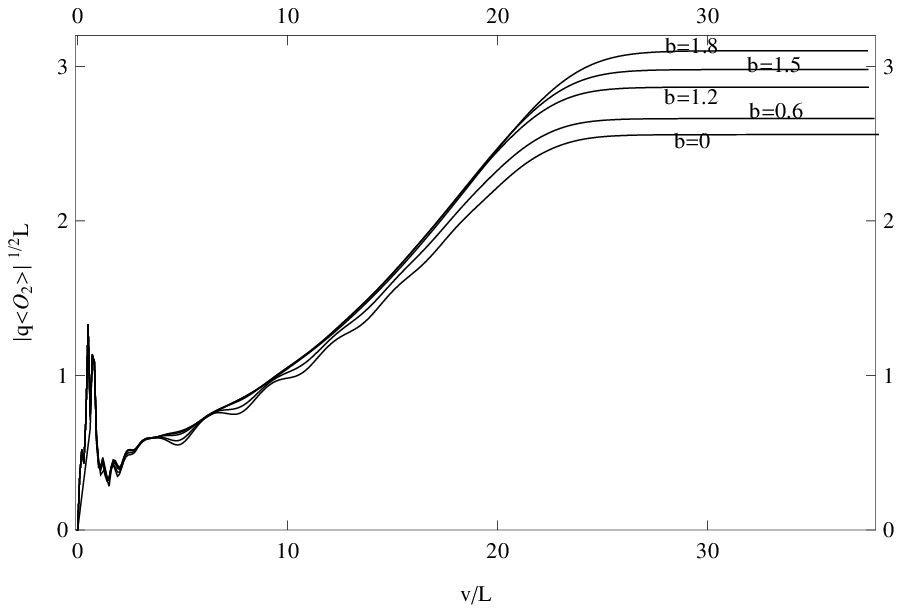}
\caption{\label{operatorsL} This figure depicts the evolution of scalar operators in the AdS boundary as a function of $v/L$ with initial temperature $T\simeq0.5T_c$ and different values of the nonlinear parameter $b$. The left panel is for a model with BINE, while the right panel is for LNE.}
\end{figure}

\begin{figure}[ht]\label{operatorsTc}
\centering
\includegraphics[width=200pt]{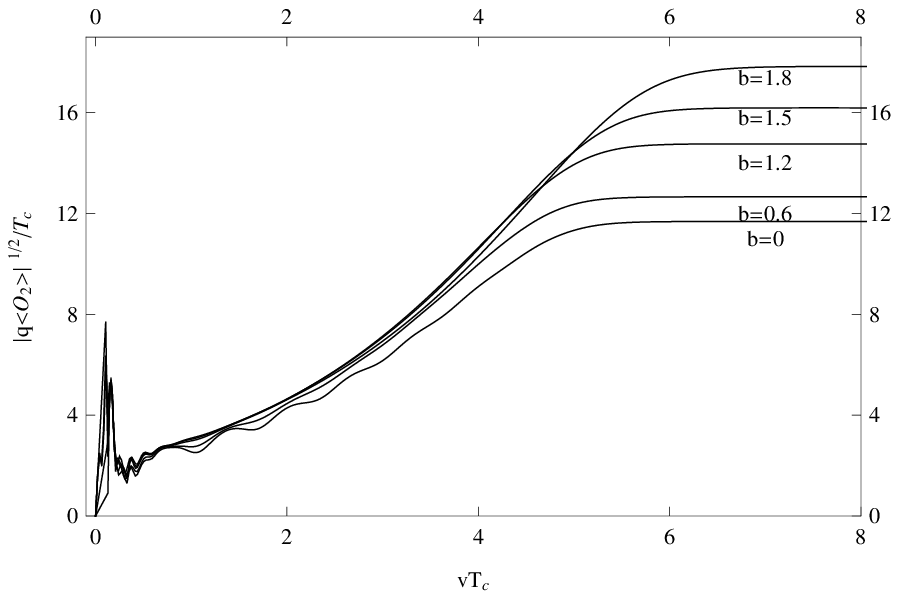}
\includegraphics[width=200pt]{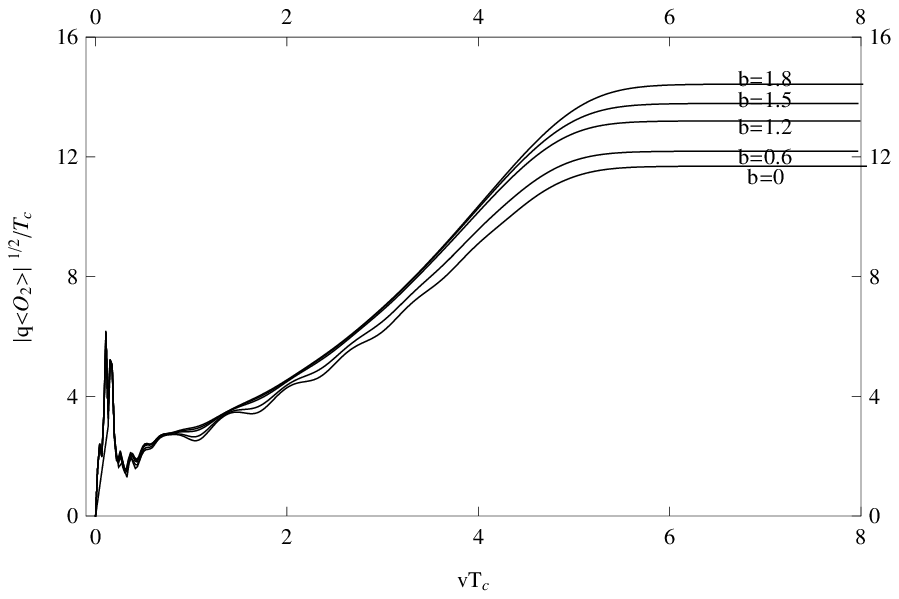}
\caption{\label{operatorsTc} This figure depicts the evolution of scalar operators in the AdS boundary as a function of $vT_c$ with initial temperature $T\simeq0.5T_c$ and different values of the nonlinear parameter $b$. The left panel is for a model with BINE, while the right panel is for LNE.}
\end{figure}

\subsection{Evolution of Horizons}
To further disclose the properties reported in the previous two subsections, we extend our discussion here to the evolution of black hole horizons. 

The apparent horizon is defined as the outer boundary of any connected trapped region. This definition makes the apparent horizon a marginally outer trapped surface on which the expansion of outgoing null geodesics vanishes. Following this definition, the apparent horizon can be obtained by
\begin{eqnarray}\label{apparent}
D\phi(v,z_{AH}(v))=0.
\end{eqnarray}

The event horizon is spanned by the outgoing light rays which neither hit the black hole singularity nor reach future null infinity. An obvious approach to locate the event horizon is to examine null geodesics \cite{Murata} by solving the equation
\begin{eqnarray}\label{event}
\dot{z}_{EH}(v)=-\frac{1}{2}F(v,z_{EH}(v)).
\end{eqnarray}

As we get the $z_{AH}(v)$ and $z_{EH}(v)$, following \cite{Murata} we can calculate the area of apparent and event horizons as
\begin{eqnarray}\label{}
Area(apparent/event~horizon)=\phi(v,z_{AH/ZH}(v))^2.\nonumber
\end{eqnarray}

\begin{figure}[ht]\label{BLNEhorizonL}
\centering
\includegraphics[width=200pt]{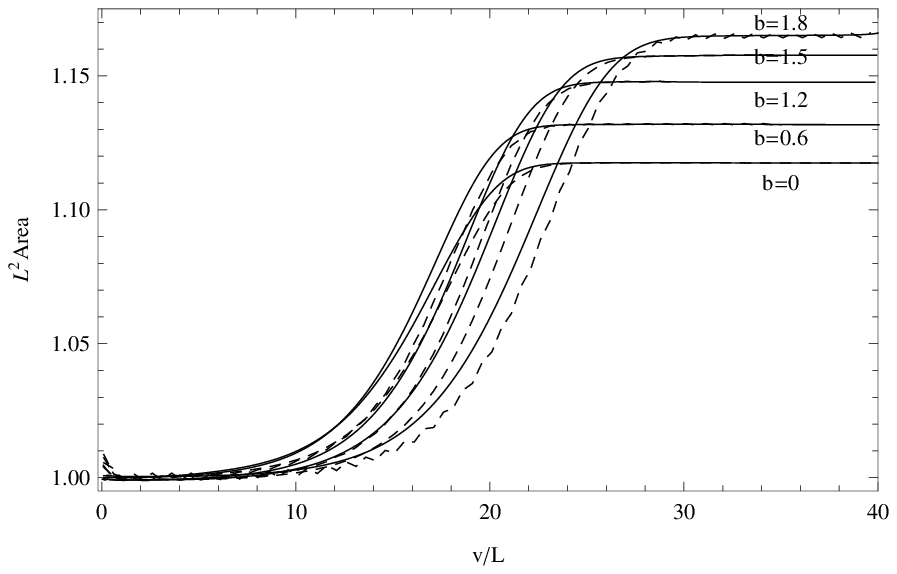}
\includegraphics[width=200pt]{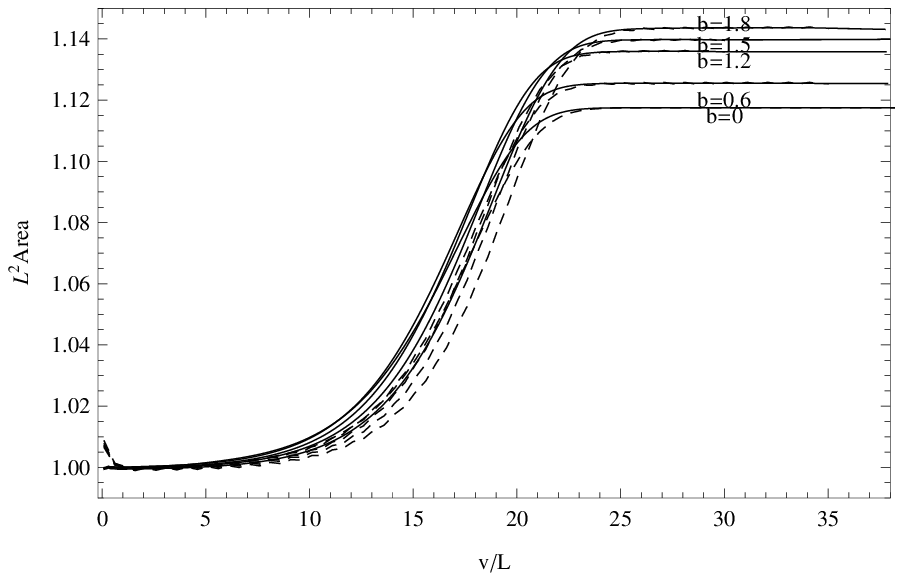}
\caption{\label{BLNEhorizon} This figure depicts the area of apparent and event horizons as functions of $v/L$ with initial temperature $T\simeq0.5T_c$ for various nonlinear parameter $b$. The left plot is for a holographic superconductor with BINE while the right one is for a model with LNE. Solid lines correspond to event horizons while dashed lines apparent horizons.}
\end{figure}

\begin{figure}[ht]\label{BINEhorizon}
\centering
\includegraphics[width=200pt]{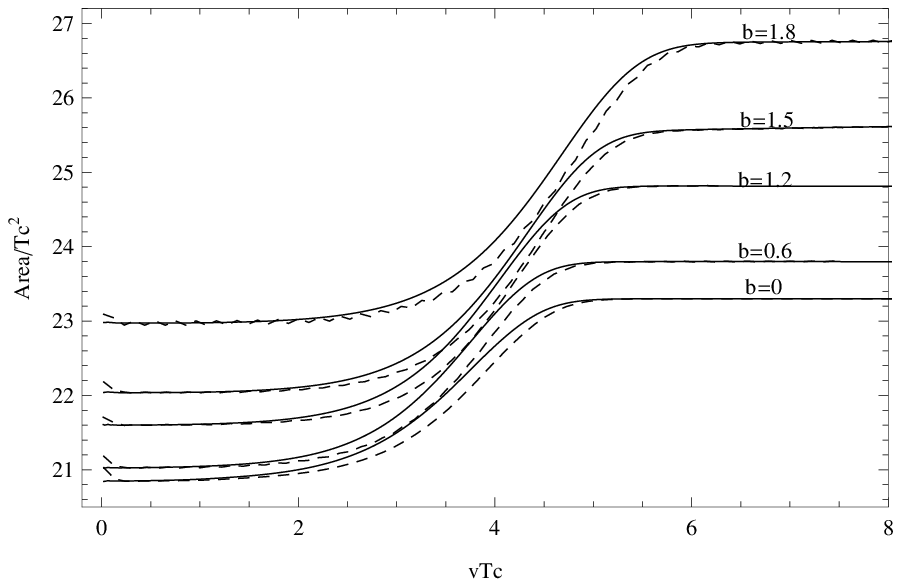}
\includegraphics[width=200pt]{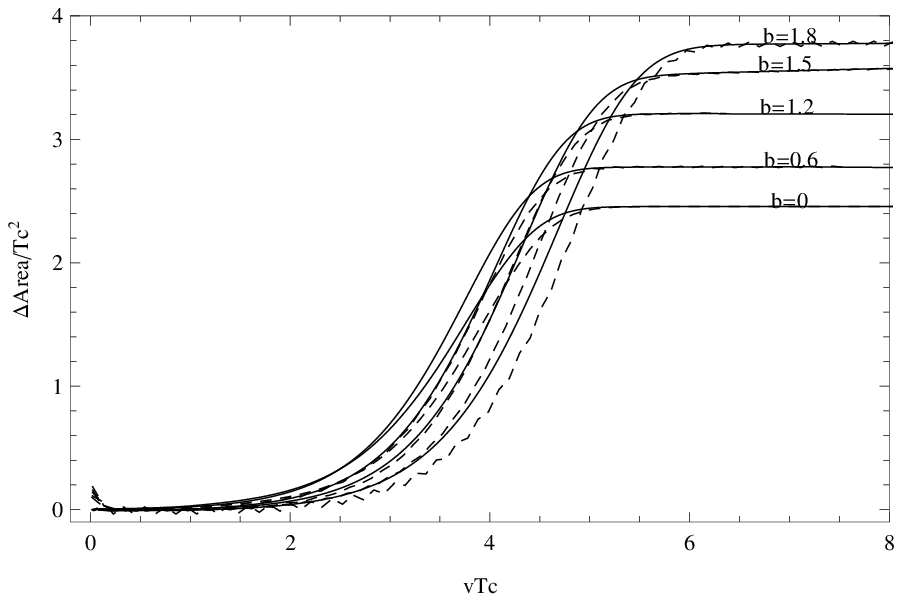}
\caption{\label{BINEhorizon} This figure depicts the area of apparent and event horizons as a function of $vT_c$ with initial temperature $T\simeq0.5T_c$ for various nonlinear parameter $b$ for a holographic superconductor with BINE. Solid lines correspond to event horizons while dashed lines apparent horizons.}
\end{figure}

\begin{figure}[ht]\label{LNEhorizon}
\centering
\includegraphics[width=200pt]{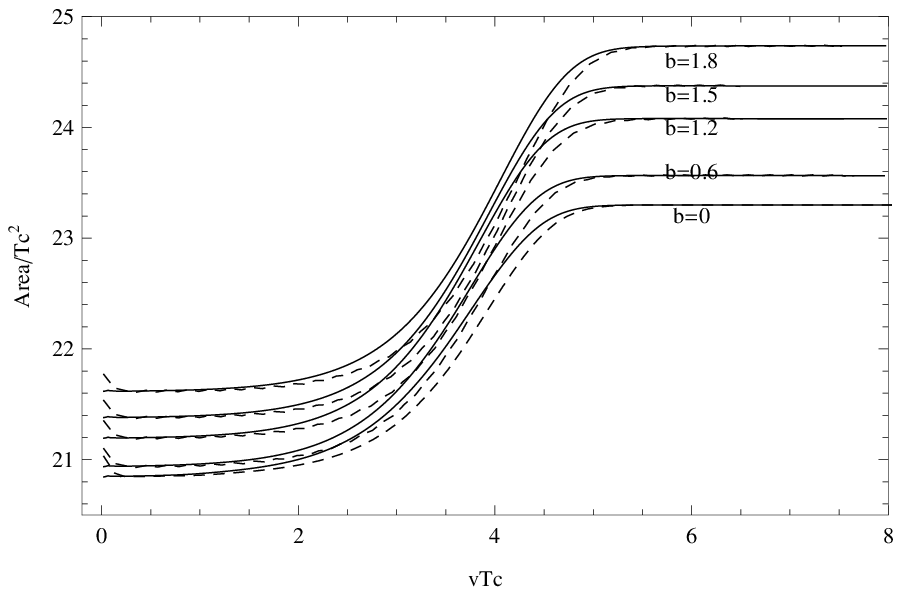}
\includegraphics[width=200pt]{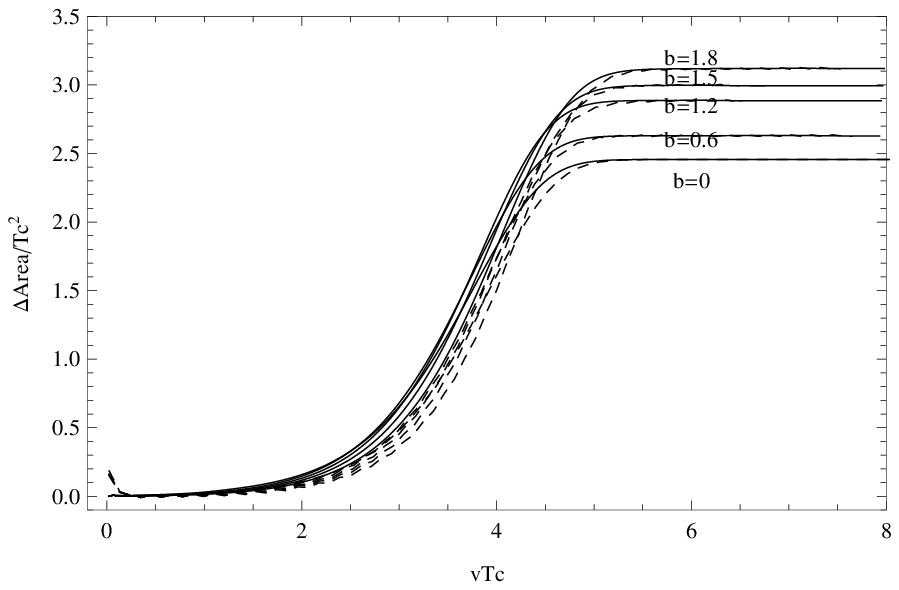}
\caption{\label{LNEhorizon} This figure depicts the area of apparent and event horizons as a function of $vT_c$ with initial temperature $T\simeq0.5T_c$ for various nonlinear parameter $b$ for a holographic superconductor with LNE. Solid lines correspond to event horizons while dashed lines apparent horizons.}
\end{figure}

Fig.\ref{BLNEhorizon} displays the time evolution of the area $L^2 Area$ of the apparent and event horizons as functions of $v/L$ in a gravitational system with nonlinear electromagnetic fields described by BINE and LNE models, respectively, where we choose the initial black hole temperature $T\simeq0.5T_c$. The initial black holes' horizon radius are set to unity, thus for an initial static AdS black hole the initial area of event horizon, which also equals the black hole entropy, coincides with that of apparent horizon. Furthermore, we can clearly see the transformation. The original stable bald AdS black hole with constant entropy evolves into the final stable hairy black hole with higher constant entropy. This transformation satisfies the second law of thermodynamics. The increasing of the area of the horizon exhibits the non-equilibrium process in the transformation. From both the two plots, it is clear that when the nonlinearity of the electromagnetic filed becomes stronger, the horizons need longer time to settle down to a constant value, which means that it needs more time to develop a stable hairy black hole. Further we see that the final stable state develops higher entropy in the gravitational system with stronger nonlinear electromagnetic field.

Figs.\ref{BINEhorizon} and \ref{LNEhorizon} displays the normalized data. The left plots present the evolution of the area $Area/{T_c}^2$. At $vT_c=0$, for the initial black holes with different $b$, the entropy no longer takes the same value, this is clearly because the impact of increasing nonlinear parameter $b$ is that the critical temperature of the system is lowered. To see more closely on the influence of the nonlinearity of the electromagnetic filed on the horizons' evolution, we plot in the right panels the net change of the horizons' area with respect to the value of the initial bald black hole, $\Delta Area/{T_c}^2$, in the transformation process. 

\subsection{Holographic Entanglement Entropy}

Finally we extend our discussion to the  holographic entanglement entropy.  The entanglement entropy is considered as an effective measure of how a given quantum system is strongly correlated. It has been proven as a powerful tool to describe different phases and the corresponding phase transitions in holographic superconductors \cite{NishiokaJHEP}-\cite{pengliu}. In this section, we will calculate the holographic entangle entropy in  the  non-equilibrium process and further disclose that the entanglement entropy can be regarded as a useful tool in describing the holographic superconductor. 

The entanglement entropy is directly related to the degrees of freedom of a system to keep track of them. From \cite{HubenyRangamani,NishiokaJHEP,BalasubramanianBernamonti}, we can see in a time-dependent background the entanglement entropy associated with a specified region on the boundary in the context of the AdS/CFT correspondence is given by the area of a co-dimension two bulk surface with vanishing expansions of null geodesics. We will follow this treatment to calculate the entanglement entropy in the time-dependent background. We use a strip geometry by considering the length of the strip, $L$, approaches infinity along the $y$-direction and stretches from $x=-h/2$ to $x=h/2$ in the $x$-direction, and the width of the strip is $h$. We express $v$ and $z$ by using parametrization $v=v(x)$ and $z=x(x)$. In the dual gravity side, this leads to the following boundary condition:
\begin{eqnarray}\label{}
v(h/2)=v(-h/2)=v_0,~~~z(h/2)=z(-h/2)=z_0,~~~v'(0)=z'(0)=0,
\end{eqnarray}
where $z_0\rightarrow 0$ is the UV cut-off and $v_0$ is the boundary time.

For the reflection symmetry of the background, $v(x)=v(-x)$ and $z(x)=z(-x)$, the resulting entanglement entropy $\bar{S}$ is given by:
\begin{eqnarray}\label{}
4\bar{S}/L=\int_{-h/2}^{h/2}\frac{dx}{z} \sqrt{z^2\phi(v,z)^2-f(v,z)v'^2-2v'z'},
\end{eqnarray}
where the prime $'$ denotes the derivative with respect to $x$, i.e. $'=\partial_x$.
This integral is independent of $x$, we have a conserved quantity
\begin{eqnarray}\label{areaeq}
\frac{z^2\phi(v,z)^4}{\phi_*^2}=z^2\phi(v,z)^2-f(v,z)v'^2-2v'z',
\end{eqnarray}
where $\phi_*$ is short for $\phi(v_*,z_*)$, $v_*=v(0)$ and $z_*=z(0)$ are constants.

From Eq. (\ref{areaeq}), we can derive two equations of motion for $v$ and $z$,
\begin{eqnarray}
&&f v'^2-1+2v'z'+z[\phi(fv'^2-2)+v''-v'^2\partial_zf/2]+z^3\phi\partial_z\phi(2+z\phi)\nonumber\\
&&-z^2[\phi^2+\phi(v'^2\partial_zf/2-v'')+(2v'z'-1)\partial_z\phi+2v'^2\partial_v\phi]=0\label{heeeom1}\\
&&(1+z\phi)(z''+v'z'\partial_zf+v'^2\partial_vf/2)-2z^2 \partial_z(z\phi)\nonumber\\
&&+z\partial_v\phi[(1+z\phi)^2-2v'z']+f[v''+\phi(zv''-2v'z')-2zv'(z'\partial_z\phi+v'\partial_v\phi)]=0\label{heeeom2}
\end{eqnarray}
Only two of Eqs. (\ref{areaeq}), (\ref{heeeom1}) and (\ref{heeeom2}) are independent, we can use two of them to solve $v(x)$ and $z(x)$. Finally, we can get the entanglement entropy density $S=\bar{S}/L$:
\begin{eqnarray}
2S(h,v_0)=\int_0^{h/2}\frac{\phi(v,z)^2}{\phi_*}dx-\text{Ln}(2/z_0).
\end{eqnarray}

In our calculation, we choose the strip width to be $1$. To show the dependence of the entanglement entropy on the parameters more clearly, we define $\Delta S= S(v_0)-S_0$, where $S_0$ is the entanglement entropy of the strip region in the normal bald AdS state. 

\begin{figure}[ht]\label{BINEHEEL}
\centering
\includegraphics[width=185pt]{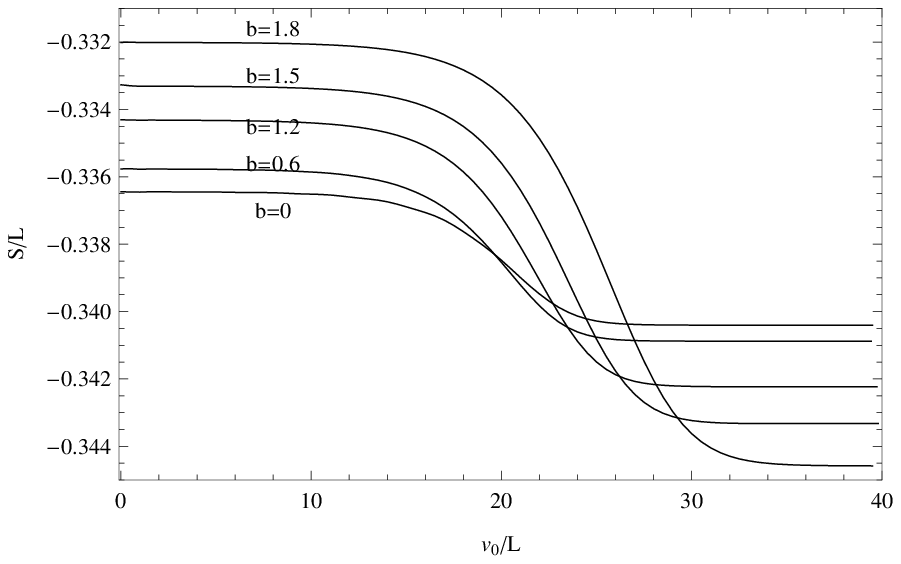}
\includegraphics[width=185pt]{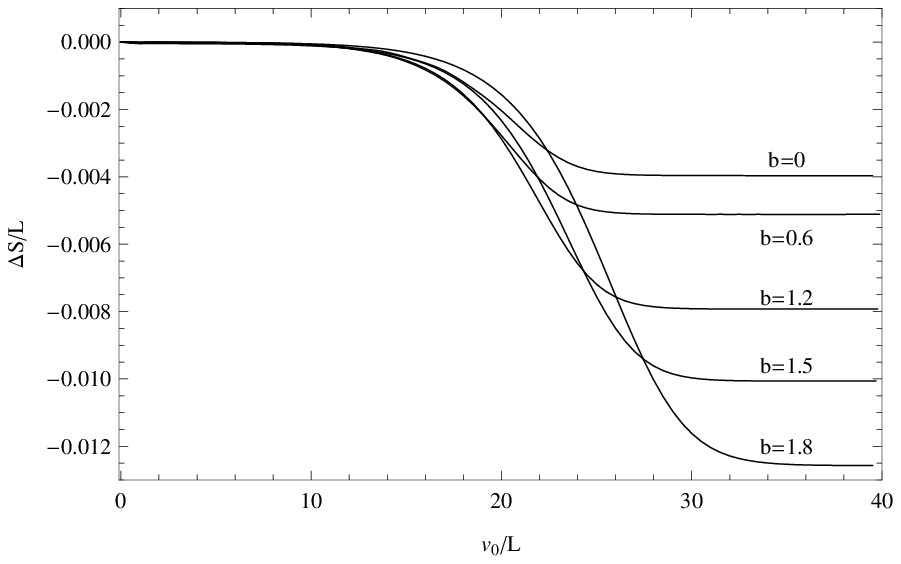}
\caption{\label{BINEHEEL} The holographic entanglement entropy is depicted as a function of $v_0/L$ with the initial tempearture $T\simeq 0.5T_c$ for various values of the nonlinear parameter $b$ in BINE.}
\end{figure}

\begin{figure}[ht]\label{LNEHEEL}
\centering
\includegraphics[width=185pt]{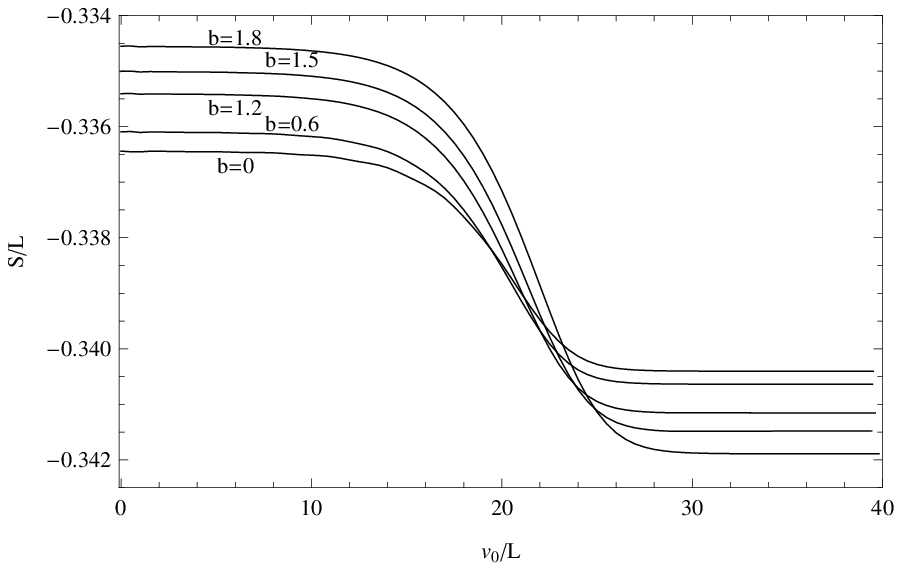}
\includegraphics[width=185pt]{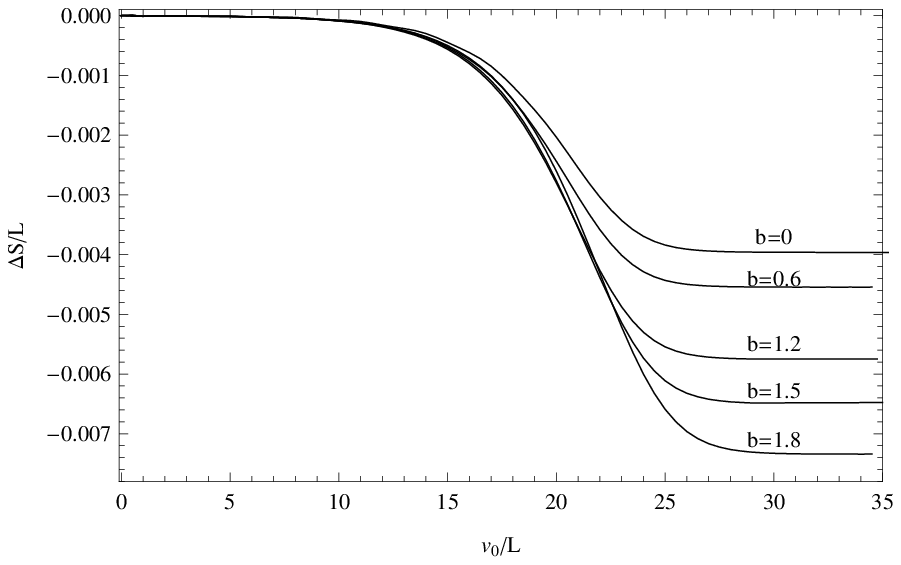}
\caption{\label{LNEHEEL}  The holographic entanglement entropy is depicted as a function of $v_0/L$ with the initial tempearture $T\simeq 0.5T_c$ for various values of the nonlinear parameter $b$ in LNE.}
\end{figure}

\begin{figure}[ht]\label{BINEHEETc}
\centering
\includegraphics[width=185pt]{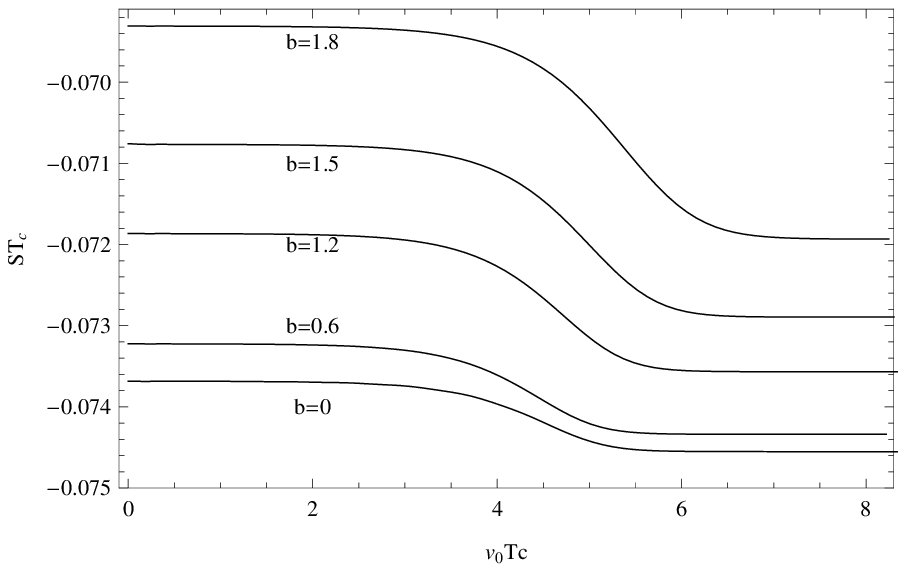}
\includegraphics[width=185pt]{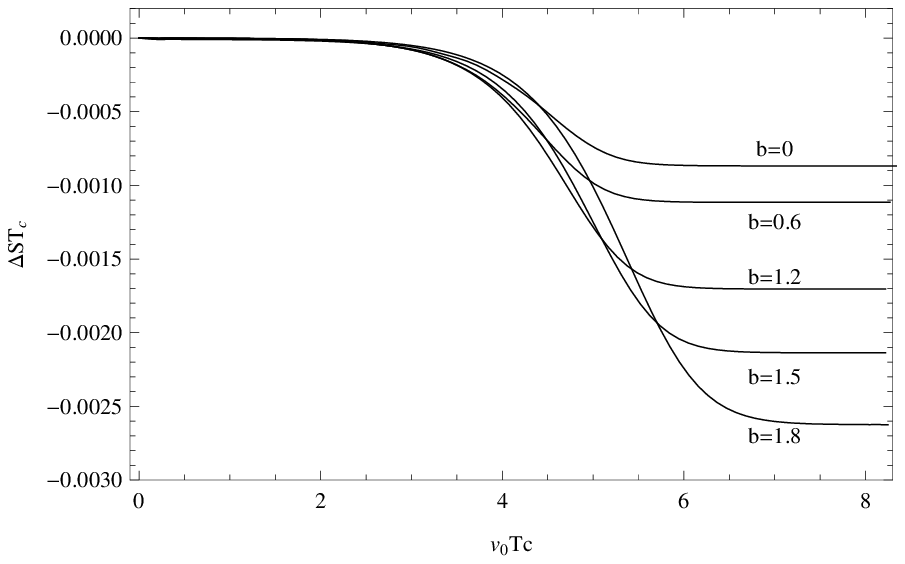}
\caption{\label{BINEHEETc} The holographic entanglement entropy is depicted as a function of $v_0T_c$ with the initial tempearture $T\simeq 0.5T_c$ for various values of the nonlinear parameter $b$ in BINE.}
\end{figure}

\begin{figure}[ht]\label{LNEHEETc}
\centering
\includegraphics[width=185pt]{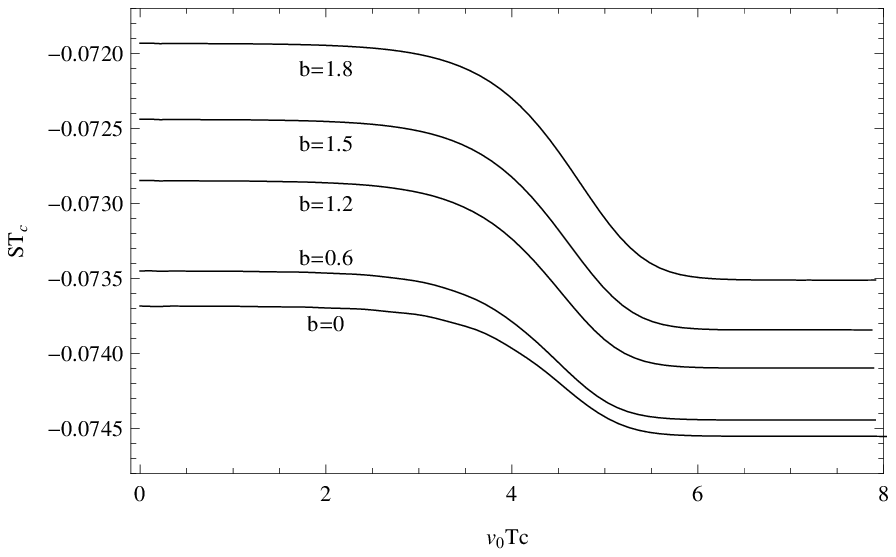}
\includegraphics[width=185pt]{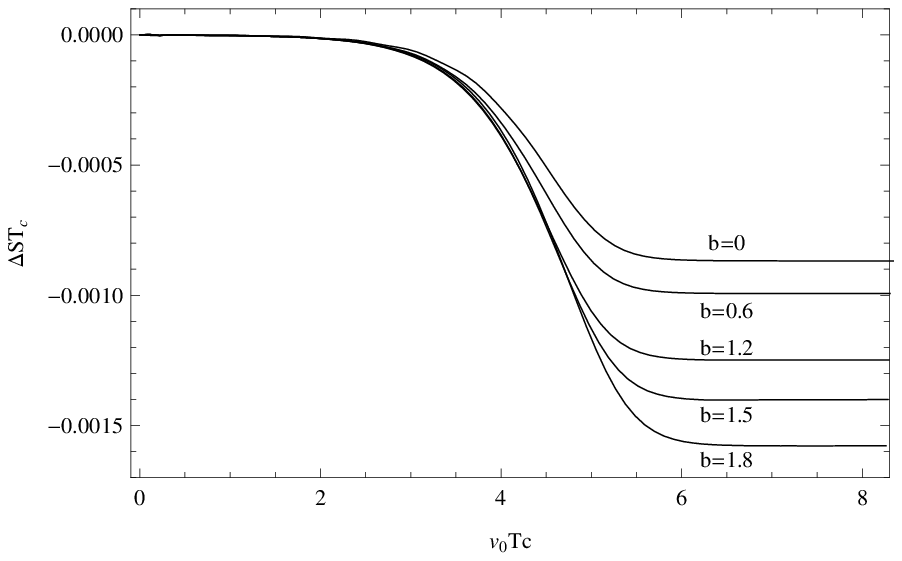}
\caption{\label{LNEHEETc}  The holographic entanglement entropy is depicted as a function of $v_0T_c$ with the initial tempearture $T\simeq 0.5T_c$ for various values of the nonlinear parameter $b$ in LNE.}
\end{figure}

Figs. \ref{BINEHEEL} and \ref{LNEHEEL} depict the raw data of holographic entanglement entropy as a function of the boundary time $v_0/L$ for various nonlinear factor $b=0,0.6,1.2,1.5$ and $1.8$ with initial temperature $T\simeq0.5T_c$ for the model of BINE and LNE, respectively. In the left plots, the curves with different $b$ intersect with each other, to see clearly the quantity change of entanglement entropy $S/L$, we plot the net change $\Delta S/L$ in the right panels. We see that below the critical temperature when the scalar field starts to condensate, the entanglement entropy becomes smaller in the non-equilibrium process. This is consistent with the expectation that in the condensation of the scalar field, certain kind of ``paring mechanics'' triggers the decrease of degrees of freedom due to the formation of ``Cooper pairs" \cite{Albashjhep1205}. When the nonlinearity of the electromagnetic field becomes stronger, the non-equilibrium process is longer and the entanglement entropy takes longer time to decrease to a constant value. This shows that, fixing $\frac{T}{T_c(b)}$, in a gravitational system with higher nonlinear electromagnetic field, it needs longer period for the system to form the ``Cooper pairs" to complete the condensate. Fixing $T/T_c(b)\simeq0.5$, the final value of the entanglement entropy for stronger nonlinearity in the electromagnetic field is lower, which shows that for the stronger nonlinear electromagnetic field there is a larger reduction in the number of degrees of freedom.

Figs. \ref{BINEHEETc} and \ref{LNEHEETc} depict the normalized holographic entanglement entropy as a function of the boundary time $v_0T_c$. In the left plots, the curves with various $b$ don't intersect with each other. This is mainly because increasing parameter $b$ the critical temperature of the system is lowered, as those shown in Table. I, thus the impact is that the curves with larger $b$ move upward much more than those with smaller $b$. In the right plots we plot the net change $\Delta S/T_c$. From figs. \ref{BINEHEETc} and \ref{LNEHEETc}, we also reach the conclusion that, fixing $\frac{T}{T_c(b)}\simeq0.5$, in a gravitational system with higher nonlinear electromagnetic field, it needs longer period for the system to complete the condensate, and for superconducting phase more degrees of freedom have condensed.

\section{Conclusions}

In this paper, we have investigated the non-equilibrium condensation process in the holographic superconductor with nonlinear electromagnetic fields. We have solved the time evolutions of the gravitational system in the bulk configurations. Following \cite{Murata,BhaseenWiseman,Bai}, we started with small scalar perturbations in the asymptotic AdS black hole, which was a static bald AdS black hole at the beginning. When the temperature is lowered below the critical value $T_c$, the initial scalar perturbations grow exponentially and the original bald AdS black hole spacetime gives way to a hairy black hole step by step. Finally the condensation of the scalar hair forces the formation of a new hairy black hole.

In this non-equilibrium process, we have observed the formation of the scalar hair step by step. Especially in the evolutions of the bulk dynamics we have found the clues why the nonlinearity in the electromagnetic fields hinders the formation of the hairy black hole. The final configuration of the gravitational system indicates that to be stable the black hole needs developing more scalar hair in the model with higher nonlinearity electromagnetic field. We also observed that, fixing $\frac{T}{T_c(b)}$, the nonlinearity makes the condensation process of the scalar field onto black hole longer. In the gravity side we also have examined the evolution of horizons in the non-equilibrium transformation process from the bald AdS black hole to the AdS hairy one. Fixing $\frac{T}{T_c(b)}$, both the evolutions of the apparent and event horizons showed that with the nonlinear corrections in the electromagnetic field, the original AdS black hole configurations require more time to finish the transformation to become hairy black hole. 

As for the superconducting order parameter in the boundary theory, we clarified for the model with stronger nonlinearity in the electromagnetic field, it costs more time for the order parameter to approach a bigger saturation value, which means the boundary system is harder to arrive at the superconducting phase. As we know, for the holographic superconductor the superconducting phase of the boundary system corresponds to a stable hairy black hole in bulk. The results obtained in both the bulk and boundary sides are consistent. 

Considering the intriguing physical meaning of the holographic entanglement entropy in studying the quantum system and its effectiveness in describing the degrees of freedom in a system, we have generalized our non-equilibrium study of the problem by examining the holographic entanglement entropy. For the stronger nonlinearity of the electromagnetic field, fixing $\frac{T}{T_c(b)}$, the non-equilibrium process is longer and the entanglement entropy takes longer time to decrease to a constant value which means it needs longer period for the boundary system to form the “Cooper pairs” to complete the condensate. It again presents us the consistent picture on the influence of the nonlinearity in the electromagnetic fields on the scalar condensation. 

\begin{acknowledgments}
Yunqi Liu thanks Shao-Jun Zhang for helpful discussion. This work was supported by the National Natural Science Foundation of China under grants No. 11505066, No. 11175270 and No. 11475065. 
\end{acknowledgments}

\end{document}